\documentclass[runningheads]{llncs}

\usepackage{graphicx}
\usepackage{bm}
\usepackage{xcolor}
\usepackage{amssymb, amsmath, amsfonts}
\usepackage{hhline, multirow, colortbl,booktabs}

\begin{document}
\title{Canonical Cortical Graph Neural Networks and its Application for Speech Enhancement in Audio-Visual Hearing Aids\thanks{The authors are grateful to FAPESP grants \#2013/07375-0, \#2014/12236-1, \#2017/02286-0, \#2018/21934-5, \#2019/07665-4, and \#2019/18287-0, CNPq grants \#307066/2017-7, and \#427968/2018-6, as well as the Engineering and Physical Sciences Research Council (EPSRC) grant EP/T021063/1.}}

\titlerunning{Canonical Cortical GNN and its Application for Audio-Visual HA}

\author{Leandro A. Passos\inst{1}\and Jo\~{a}o Paulo Papa\inst{2} \and Amir Hussain\inst{3} \and Ahsan Adeel\inst{1,4}}

\authorrunning{Passos et al.}

\institute{CMI Lab, School of Engineering and Informatics, University of Wolverhampton, Wolverhampton, United Kingdom
\and Department of Computing, S\~ao Paulo State University, Bauru, Brazil 
\and School of Computing, Edinburgh Napier University, Edinburgh, Scotland, United Kingdom.\\
\and deepCI.org 20/1 Parkside Terrace, Edinburgh, United Kingdom\\
\email{ahsan.adeel@deepci.org} }
\maketitle              
\begin{abstract}

Despite the recent success of machine learning algorithms, most models face drawbacks when considering more complex tasks requiring interaction between different sources, such as multimodal input data and logical time sequences. On the other hand, the biological brain is highly sharpened in this sense, empowered to automatically manage and integrate such streams of information. In this context, this work draws inspiration from recent discoveries in brain cortical circuits to propose a more biologically plausible self-supervised machine learning approach. This combines multimodal information using intra-layer modulations together with Canonical Correlation Analysis, and a memory mechanism to keep track of temporal data, the overall approach termed Canonical Cortical Graph Neural networks. This is shown to outperform recent state-of-the-art models in terms of clean audio reconstruction and energy efficiency for a benchmark audio-visual speech dataset. The enhanced performance is demonstrated through a reduced and smother neuron firing rate distribution. suggesting that the proposed model is amenable for speech enhancement in future audio-visual hearing aid devices.

\end{abstract}

\section{Introduction}
\label{s.intro}

According to the World Health Organization (WHO), the number of people requiring hearing rehabilitation in the world is estimated to rise from $430$ million nowadays up to $700$ million until 2050, with nearly $2.5$ billion people presenting at least some degree of hearing impairment~\cite{world2021hearing}. Despite the impairment itself, deafness also impacts on social relationships~\cite{kramer1995factors,huang2021hearing} and environment perception~\cite{noble1998self}, leading to other psychological and health conditions~\cite{helvik2006psychological}. Over such circumstances, employing high-end energy-efficient technological approaches to build cross-modal sensory devices, i.e., combining audio and visual inputs to enhance hearing aid devices, seems a plausible way to improve individuals' life quality.

\begin{sloppypar}
In the last decades, machine learning-based techniques have shown themselves as a suitable approach to tackle issues related to virtually any field of science, industry, or even daily life, ranging from computer vision~\cite{zeng2022small} to medicine~\cite{de2021computer}, and satellite imagery processing~\cite{da2020critical,santos2021ddipnet}. Machine learning also has been successfully employed in the context of speech enhancement~\cite{xu2014regression,yu2022setransformer}, whose aim is to enhance speech quality and intelligibility when noise degrades them significantly~\cite{adeel2019lip}. Nevertheless, such methods may suffer massive performance degradation in the presence of overwhelming noise~\cite{benesty2011speech}. Many works address this problem using multimodal audio-visual (AV) information fusion. Combining AV information usually demands more sophisticated approaches, which intrinsically comprises several challenges, like data alignment, finding semantic gaps between low-level features and high-level information~\cite{bokhari2013multimodal}, and learning coherent and correlated latent patterns on different input modalities.
\end{sloppypar}

Combining noisy audio and clean images for clean signal reconstruction is analogous to reading the lips and body movements of a speaker in a boisterous environment, e.g., a pub with loud music, to obtain some additional information and create a context to enhance the information quality of a speech suppressed by the loud sound. Ngiam et at.~\cite{ngiam2011multimodal}, for instance, proposed a multimodal method capable of improving a target modality feature representation. Further, Adeel et al.~\cite{adeel2018real} provided several improvements to the field, presenting a chaotic model for lip-reading integrating Internet of Things (IoT) and 5G Cloud-Radio Access Network, further improving AV information transmission for real-time speech reconstruction. Further work employ deep learning-based approach to exploits AV cues to estimate clean audio~\cite{adeel2020contextual,kumar2022deep}, also considering distinct language speakers~\cite{gogate2020cochleanet}.

Recently, Passos et al.~\cite{passos2022} proposed a multimodal self-supervised Graph Neural Network (GNN) that combines AV data through using Canonical Correlation Analysis Graph Neural Networks (CCA-GNN)~\cite{zhang2021canonical}, also modeling the temporal information in the graph using the so-called prior-frame positional encoding. The method obtained outstanding results considering audio reconstruction and energy efficiency, analyzed in terms of neuronal activation rate.

Despite the advantages presented~\cite{passos2022}, the model lacks some points in the context of a biologically plausible approach. In this context, Passos et al.~\cite{passos2022multimodal} proposed a multimodal approach using burst-dependent learning~\cite{payeur2021burst}, a method inspired by more recent studies on the physiological mechanism of pyramidal neurons that regulates the learning by the frequency of bursts, where the credit assignment problem is addressed by the primary principles of pyramidal neurons suggested by K\"{o}rding and K\"{o}nig~\cite{kording2001supervised}. In parallel, the study of canonical cortical circuits~\cite{canonicalCortical,canonicalLaminar} provides some interesting insights regarding the brain procedure toward multimodal information processing. Canonical cortical circuits model pyramidal neurons to receive different modalities of information, modulated in an excitatory or inhibitory fashion on deeper layers. Moreover, biologically plausible models should not underestimate the importance of memory in the learning process, which performs a fundamental role in information inference, acting as an intrinsic context for novelty comprehension.

The attributes mentioned above motivated the development of the Canonical Cortical Graph Neural Network, a novel self-supervised architecture that remodels and improves the ideas developed in~\cite{passos2022} by introducing a more biologically plausible approach to modulating and filtering the multimodal information inside the so-called cortical graph layers. It also introduces a memory concept inside each cortical graph block, composed of a mechanism to ``forget'' irrelevant facts and update itself considering every consecutive node, which is presented in a logical time-step sequence using prior-frame-based positional encoding. Experiments conducted over the AV ChiME3~\cite{adeel2019lip} dataset show the Canonical Cortical Graph Neural Network obtained state-of-the-art results, outperforming CCA-GNN in the contexts of faster convergence, audio reconstruction, and firing neuron activation rates. Regarding the latter, the model not only obtained lower rates of activation, but also produced smothier firing-rate distributions.

The main contributions of this paper are presented as follows:

\begin{enumerate}
	\item To propose the Canonical Cortical Graph Neural Network, a biologically plausible model for multimodal information feature extraction.
	\item To introduce a novel paradigm for intra-layer information fusion and memory modeling.
	\item To provide a self-supervised energy-efficient model for correlated feature extraction for AV-based clean audio data reconstruction.
	\item To foster the literature regarding speech enhancement and AV hearing aids.
\end{enumerate}

The remainder of this paper is presented as follows. Section~\ref{s.background} provides a theoretical background regarding Graph Neural Networks with Canonical Correlation Analysis and the prior frame-based graph positional encoding, while Section~\ref{s.proposed} introduces the novel Canonical Cortical Graph Neural Network. Further, Sections~\ref{s.methodology} and~\ref{s.experiments} present the methodology and the experimental results, respectively. Finally, Section~\ref{s.conclusion} states  conclusions and future works.

\section{Theoretical Background}
\label{s.background}

This section provides a brief theoretical background regarding Graph Neural Networks with Canonical Correlation Analysis and the prior frame-based graph positional enconding.

\subsection{Graph Neural Networks with Canonical Correlation Analysis}
\label{ss.cca_gnn}

Let $G = (X, A)$ be a graph where $A \in \mathbb{R}^{N \times N}$ represents the adjacency matrix and $X\in \mathbb{R}^{N\times F}$ the input data represented by graph nodes. Additionally, $F$ represents the feature space dimension, and $N$ denotes the number of nodes. The CCA-GNN~\cite{zhang2021canonical} comprises three main steps, i.e., a random graph generator $T$,  a graph neural network encoder $f_\theta$, where $\theta$ denotes the network's learnable weights, and an objective function based on Canonical Correlation Analysis. The graph generator produces two augmented versions of the same graph, which are presented to the graph neural network encoder for further computing and maximizing the canonical correlation between their outputs.

\begin{figure*}[!ht]
  \centerline{
    \begin{tabular}{c}
	\includegraphics[width=\textwidth]{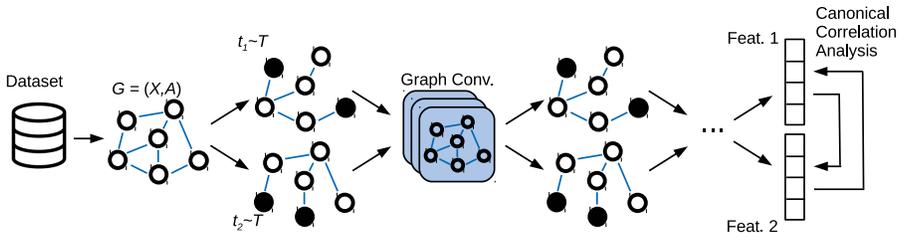} 
    \end{tabular}}
    \caption{Canonical Correlation Analysis Graph Neural Network. Each sample represents a node in a graph whose edges describe the relationship between pairs of samples. The random graph generator $T$ produces two augmented versions of this graph, which are employed to feed the GNN model. The output of both versions is compared using the canonical correlation analysis, and the network parameters are adjusted to maximize this metric.}
  \label{f.CCAGNN}
\end{figure*}

The idea behind such an approach is discarding decorrelated components while preserving correlated ones. In a nutshell, the model tries to keep the more significant information present in both augmented versions and to avoid individual behaviors, such as anomalies and noise. Figure~\ref{f.CCAGNN} depicts the Canonical Correlation Analysis Graph Neural Network.

The graph augmentation process comprises the same approach presented in~\cite{zhu2020deep,thakoor2021bootstrapped}, which conducts a random feature masking and edge dropping. In this context, each $t_i\sim T$ comprises a distinct view, i.e., a transformed version of $G$, sampled at each iteration $i$.

The encoder is composed of a two-layered GNN but can be easily replaceable by any fancier model. The target function is designed to model the learning process as a canonical correlation maximization problem~\cite{chang2018scalable} using a self-supervised approach that considers two normalized views, $\bm{Z}_A$ and $\bm{Z}_B$, produced over randomly augmented versions of the original graph. The objective is to maximize the correlation between these views, formalized as follows:

\begin{equation}
{\cal L}(\bm{Z}_A, \bm{Z}_B) = ||\bm{Z}_A - \bm{Z}_B||_{F}^2+\lambda\left(||\bm{Z}_A^T\bm{Z}_A-\bm{I}||_F^2+||\bm{Z}_B^T\bm{Z}_B-\bm{I}||_F^2\right), 
\label{e.canonicalCorrelation}
\end{equation}
where $\lambda$ is a non-negative trading-off hyperparameter, and $I$ is the identity matrix. The left term indicates the invariance term, which is responsible for minimizing the invariance between the two views. In contrast, the term on the right side describes the decorrelation term, which facilitates distinct features to capture different semantics through a regularization procedure.

The terms in Equation~\ref{e.canonicalCorrelation} can be decomposed using a variance-covariance perspective~\cite{tian2021understanding}. Let $\bm{s}$ be an augmented version of the graph sampled from an input $\bm{x}$, and $\bm{Z_s}$ is a view of $\bm{s}$ obtained from a decoder output. The invariance term is minimized using expectation, described as follows:

\begin{equation}
{\cal L}_{inv} = ||\bm{Z}_A - \bm{Z}_B||_{F}^2= \sum_{i=1}^N\sum_{k=1}^D(z_{i,j}^A-z_{i,j}^B)^2 \cong \mathbb{E}_{\bm{x}}\left[\sum_{k=1}^D\mathbb{V}_{\bm{s}|\bm{x}}[\bm{Z_s},k]\right]*2N,
\label{e.loss_inv}
\end{equation}
where $\mathbb{V}$ denotes the variance. Similarly, one can formalize the decorrelation term as follows:

\begin{equation}
{\cal L}_{dec} = ||\bm{Z}_S^T\bm{Z}_S-\bm{I}||_F^2= ||\bm{Cov}_{\bm{s}}[\bm{z}]-I||_F^2
 \cong \sum_{i\neq j}(\rho_{i,j}^{\bm{z_s}})^2,\forall \bm{Z}_S\in\{\bm{Z}_A,\bm{Z}_B\},
\label{e.loss_dec}
\end{equation}
where  $\rho$ denotes the Pearson correlation coefficient, and $\bm{Cov}$ stands for the covariance matrix.

\subsection{Prior Frame-based Graph Positional Enconding}
\label{ss.temporalRelationship}

\begin{figure}[!htb]
  \centerline{
    \begin{tabular}{c}
	\includegraphics[width=.7\textwidth]{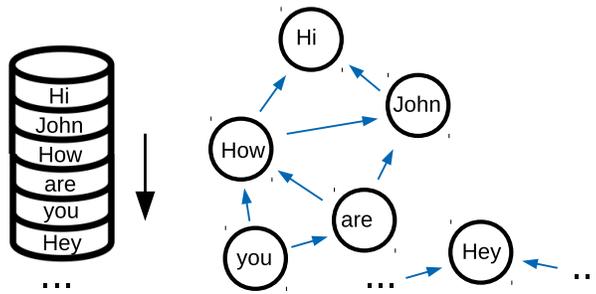} 
    \end{tabular}}

    \caption{Prior frame-based node neighborhood modeling considering $2$ prior frames.}
  \label{f.sequentialGraph}
\end{figure}

A time-sequence based approach for graph positional encoding was recently proposed by Passos et al.~\cite{passos2022}. The method computes the node neighborhood by connecting them to their $k$ prior frame nodes in time sequence and attributing connection weights to the edges according to their distances in this time-based space. Figure~\ref{f.sequentialGraph} depicts this idea.

The calculation of the edge weight $w_{ij}$ that connects a node $i$ to a node $j$ is performed as follows:

\begin{equation}
w_{ij}=k+1 - d_{ij},
\label{e.weight_distribution}
\end{equation}
where $d_{ij}$ denotes the distance from node $i$ to  node $j$ in a frame-step space. Those values are stored in a distance matrix used to compute the positional encoding of the nodes.

\section{Canonical Cortical Graph Neural Network}
\label{s.proposed}

\begin{figure*}[!htb]
  \centerline{
    \begin{tabular}{c}
	\includegraphics[width=\textwidth]{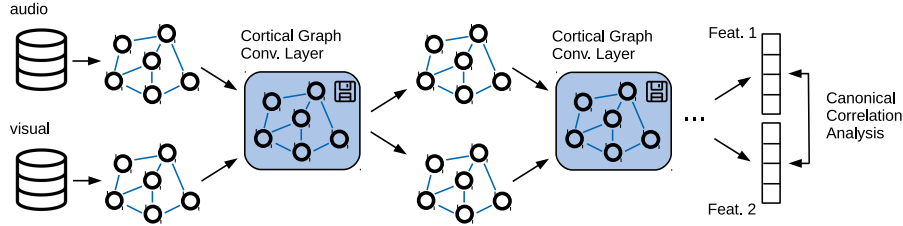} 
    \end{tabular}}
    \caption{Canonical Cortical  Graph Neural Network.}
  \label{f.corticalCanonicalGraph}
\end{figure*}

\begin{figure*}[!htb]
  \centerline{
    \begin{tabular}{c}
	\includegraphics[width=\textwidth]{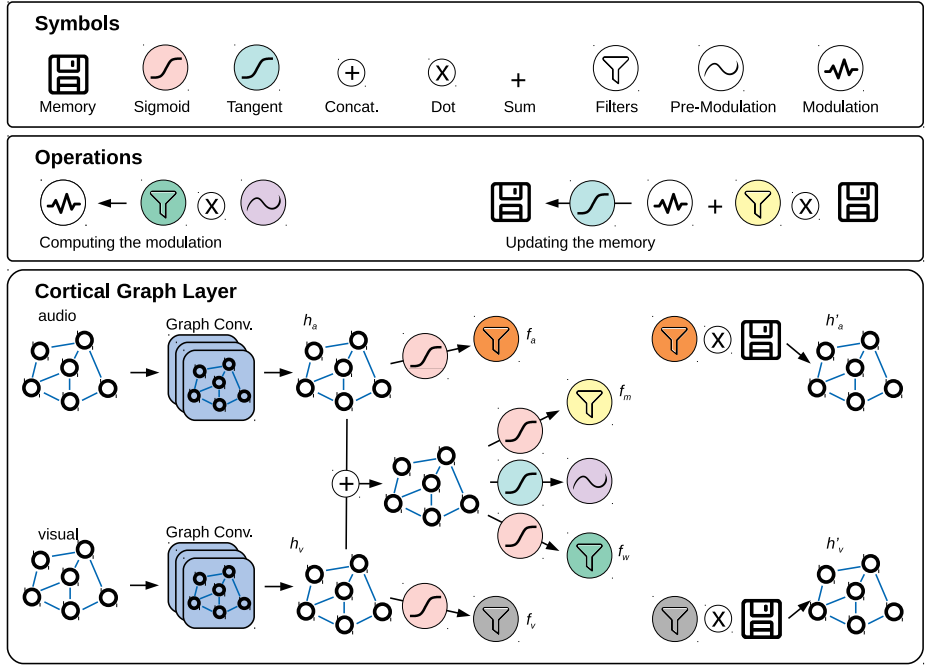} 
    \end{tabular}}
    \caption{Cortical graph layer. The top and middle frames describe the symbols and operations employed in the cortical graph block, respectively. The bottom block depicts the layer's pipeline.}
  \label{f.corticalLayer}
\end{figure*}

This paper presents a novel self-supervised approach for training multi-modal graph neural networks in a more biologically plausible way. In this context, the architecture combines several concepts inspired on cortical circuits observed in the brain~\cite{canonicalLaminar,canonicalCortical} to model memory and multimodal information fusion with canonical correlation analysis~\cite{chang2018scalable}, and to maximize the correlation of the information extracted from different inputs. Figure~\ref{f.corticalCanonicalGraph} presents a general overview of the proposed model, which suggests the more interesting procedures of the model are implemented at layer level. Therefore, Figure~\ref{f.corticalLayer} illustrates this in-layer process, also depicting the behavior of each operation.

A formal description of the procedures depicted in Figure~\ref{f.corticalLayer} and performed inside each cortical layer is provided bellow. Firstly, one should compute the audio $f_{a}$, visual $f_{v}$, memory $f_{m}$, and modulation $f_{w}$ filters as follow:

\begin{equation}
	\label{e.f_a}
	f_{a} = \sigma\left({\bm{W}_ah_a+b_a}\right),
\end{equation}

\begin{equation}
	\label{e.f_v}
	f_{v} = \sigma\left({\bm{W}_vh_v+b_v}\right),
\end{equation}

\begin{equation}
	\label{e.f_m}
	f_{m} = \sigma\left({\bm{W}_m[h_a,h_v]+b_m}\right),
\end{equation}
and

\begin{equation}
	\label{e.f_w}
	f_{w} = \sigma\left({\bm{W}_w[h_a,h_v]+b_w}\right),
\end{equation}
where $\sigma$ stands for the Sigmoid function, $\bm{W}_a$, $\bm{W}_v$, $\bm{W}_m$, and $\bm{W}_w$, are the weight matrices for the audio, visual, memory, and modulation filters, respectively, and $b_a$, $b_v$, $b_m$, and $b_w$ are the biases for the audio, visual, memory, and modulation filters, respectively, while $[h_a,h_v]$ denotes the concatenation of the audio $h_a$ and the visual $h_v$ graph convolution outputs. Further, the pre-modulation $\rho$ is computed as follows:

\begin{equation}
	\label{e.rho}
	\rho = tanh\left({\bm{W}_{\rho}[h_a,h_v]+b_{\rho}}\right),
\end{equation}
where $\bm{W}_{\rho}$ and $b_{\rho}$ are the pre-modulation weight matrix and bias, respectively. Finally, the modulation $\omega$ is computed as follows:

\begin{equation}
	\label{e.omega}
	\omega = f_{w} \otimes \rho,
\end{equation}
where $\otimes$ stands for the dot product.

The memory $\mu$ is the more tricky updating since it considers both ``forgetting''\ irrelevant memories using the memory filter $f_m$ and introducing new experiences presented in the modulation $\omega$. Moreover, the operation is performed individually for each node $n\in \{0,\dots,N\}$ since they are connected in a logical temporal sequence established by the prior frame positional encoding~\cite{passos2022}, described as follows:

\begin{equation}
	\label{e.mu_n}
	\mu^n = \omega^n + \left(f_m^n\otimes \mu^{n-1}\right),
\end{equation}
where $\mu^n$, $\omega^n$, and $f_m^n$ are the node's $n$ memory, modulation, and memory filter. In the sequence, the memory is updated as follows:

\begin{equation}
	\label{e.mu}
	\mu = tanh\left({\bm{W}_{\mu}\mu+b_{\mu}}\right),
\end{equation}
where $\bm{W}_{\mu}$ and $b_{\mu}$ are the memory weight matrix and bias, respectively. Finally, the layer output, i.e., the new node representation of the audio and visual graphs $h_a'$ and $h_v'$, respectively, are computed as follows:

\begin{equation}
	\label{e.h_a_line}
	h_a' = \mu \otimes f_{a}
\end{equation}
and
\begin{equation}
	\label{e.h_v_line}
	h_v' = \mu \otimes f_{v}.
\end{equation}

Finally, $h_a'$ and $h_v'$ become the the node representations of the subsequent layer audio and visual graphs, respectively, in case of an intermediate layer. Regarding the output layer, the model performs the canonical correlation analysis between $h_a'$ and $h_v'$ using Equation~\ref{e.canonicalCorrelation} and backpropagates this value to optimize the network parameters. 

\section{Methodology}
\label{s.methodology}

This section describes the dataset and configuration employed during the experiments.

\subsection{AV ChiME3 Dataset}
\label{ss.dataset}

The dataset used in this paper aims to combine environmental information, i.e., audio and visuals, to train an efficient model for enhancing and amplifying clean audio signals considering multimodal data. The dataset comprises triples composed of image, clean audio, and noisy audio signals, from which the image and noisy audio denote the model's input while the clean signal stands for the desired output, i.e., the instance target in the context of supervised learning. The videos are extracted from Grid~\cite{cooke2006audio} dataset, in which different classes of noises (public transport,  pedestrian area, street junction, cafe) with signal to noise ratios (SNR) ranging from -12 to 12dB extracted from ChiME3~\cite{barker2015third} are introduced, composing the AV ChiME3~\cite{adeel2019lip} dataset. Further, the samples are preprocessed to improve the sentence alignment and incorporate multiple visual frames to include temporal data. In total, the dataset contains $989$ sequences from $5$ different speakers, described as one black male, two white males, and two white females. Each sequence comprises $48$ frames, summing up to a total of $47,472$ synchronized triples of samples.

\subsubsection{Audio feature extraction}
\label{sss.audioDataset}

\begin{sloppypar}
Log-FB vectors were employed to extract both clean and noisy audio features. The technique samples the audio signal at $22,050$kHz for further segmenting it into $M$ $16$ms frames with $800$ samples per frame and $62.5\%$ increment rate. Furthermore, it uses a hamming window and Fourier transformations to produce a 2048-bin power spectrum. Moreover, it employs a logarithmic compression to obtain the $22$-dimensional log-FB signals.
\end{sloppypar}

\subsubsection{Visual feature extraction}
\label{sss.visualDataset}

The samples extracted to generate the visual set of features were obtained using an encoder-decoder architecture over the Grid Corpus dataset. Lip-regions were detected using Viola-Jones~\cite{viola2001rapid} algorithm, for further tracking the frame sequence using the method proposed in~\cite{ross2008incremental}. Additionally, a manual effort was employed to randomly inspect the sentences, ensuring good lip tracking~\cite{abel2016data}. The encoder-decoder approach is then used to create vectors of pixel intensities, in which the $50$ first components are vectorized in a zigzag order and then interpolated to match the equivalent audio sequence.

\subsection{Experimental Setup}
\label{ss.setup}

The experiments conducted in this paper compare the proposed Canonical Cortical Graph Neural Networks against the recent state-of-the-art model CCA-GNN for the task of multimodal clean audio reconstruction based on noisy audio and visual features, extracted using logFB and an encoder-decoder approach, respectively, as described in Section~\ref{ss.dataset}. Both models comprise a similar architecture composed of two hidden layers, the first comprising $512$ and the second $256$ neurons. The parameters were selected empirically. The learning is conducted to maximize the canonical correlation analysis for coherent feature extraction during $200$ epochs using the Adam optimizer with a learning rate of $10^{-3}$ and a trading-off parameter of $\lambda=0.0001$ (Equation~\ref{e.canonicalCorrelation}). At the same time, the hyperparameters of the CCA-GNN follow the configuration employed in~\cite{passos2022}. Finally, the graphs are generated considering eight distinct prior-frame scenarios, i.e., $k\in[3, 5, 7, 10, 15, 20, 25, 30]$. Notice that the plots provided in experimental results comprise only $k\in[3,10,30]$ to illustrate the difference between a low, medium, and a high number of neighbors/prior frames.

After maximizing the canonical correlation analysis between the two channels, the features extracted by the networks are employed to feed a dense layer responsible for reconstructing the clean signal by minimizing the mean squared error as the cost function. The model is optimized during $2,000$ epochs using the Adam optimizer with a learning rate of $0.005$ and a weight decay of $0.0004$.

The dataset was divided into $20$ folds to provide an in-depth statistical analysis. Each fold comprises $50$ sequences of $48$ frames each, summing up to $2,400$ samples per fold. As stated in Section~\ref{ss.dataset}, the dataset is formed by three subsets: (i) clean audio, (ii) noisy audio, and (iii) clean visual. The noisy audio and the clean visual input are used to feed the multimodal GNNs, while the clean audio is considered the reconstruction target. Finally, each fold is split into train, validation, and test sets, following the proportions of $60\%$, $20\%$, and $20\%$, respectively. The Wilcoxon signed-rank test~\cite{Wilcoxon:45} with $5\%$ of significance was considered for statistical evaluation.

\section{Experiments}
\label{s.experiments}

This section exploits the superiority of the proposed Canonical Cortical Graph Neural Networks over the state-of-the-art CCA-GNN. The experimental results consider the contexts of feature extraction analysis, clean audio signal reconstruction, as well as neuronal activation and energy efficiency.

\subsection{Feature Extraction Analysis}
\label{ss.featureExtraction}

\begin{figure}[!htb]
  \centerline{
    \begin{tabular}{cc}
	\includegraphics[width=0.9\textwidth]{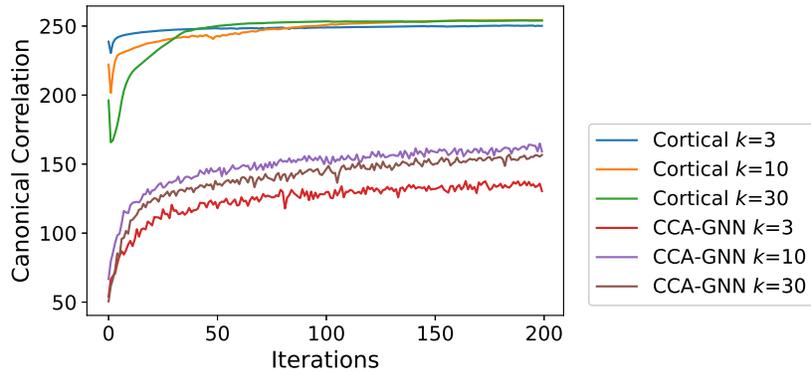}
    \end{tabular}}
    \caption{Canonical correlation analisys maximization. }
  \label{f.feature_extraction}
\end{figure}

This section explores the task of self-supervised feature extraction in terms of canonical correlation analysis. The idea consists of extracting correlated features considering both the audio and visual channels, contributing to better features for clean audio reconstruction. In this context, the Canonical Cortical Graph Neural Networks, presented in Figure~\ref{f.feature_extraction} as Cortical, showed a performance $75\%$ higher than the CCA-GNN, on average, considering a small, medium, and high neighborhood, i.e., $k\in[3,10,30]$.

\subsection{Clean Signal Reconstruction}
\label{ss.reconstruction}

Figure~\ref{f.reconstruction_convergence} provides some insights regarding the convergence of the model regarding the Mean Squared Error (MSE) over the testing set reconstruction. In this context, one can observe that (i) the proposed model (Cortical) converges a bit faster than CCA-GNN (the blue line when $k=3$) and remain the best result until completing the $2,000$ epochs; (ii) the convergence is faster for small values of $k$, i.e., $k=3$, for both Cortical and CCA-GNN; (iii) the proposed model presents an overfitting behavior for larger values of $k$, described by the orange ($k=10$) and green ($k=30$) lines.

\begin{figure}[!htb]
  \centerline{
    \begin{tabular}{c}
	\includegraphics[width=0.9\textwidth]{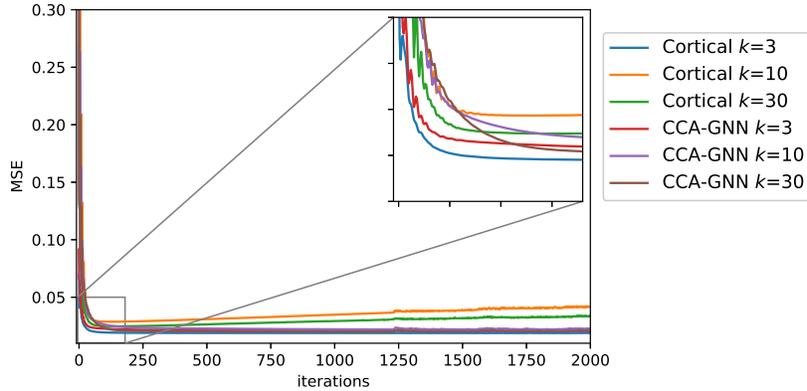} 
    \end{tabular}}30
    \caption{Clean audio reconstruction error convergence considering the proposed model (Cortical) and CCA-GNN architectures based on the noisy audio and clean visuals.}
  \label{f.reconstruction_convergence}
\end{figure}

Further, Table~\ref{t.reconstruction_audioVisual_2} provides the final MSE values over the testing set considering eight distinct $k$ scenarios, i.e., $k\in[3,5,7,10,15,20,25,30]$. Notice bold values stand for the best values considering the Wilcoxon signed-rank test~\cite{Wilcoxon:45} with a significance of $5\%$. Such results show that the proposed model presents a better behavior when exposed to a reduced number of neighbors, i.e., the historical information is reduced, obtaining the lowest MSE overall over this scenario. This result can be explained by the memory implemented in the architecture, i.e., since the memory tries to model the predictions based on past frames, a longer temporal sequence makes this information ambiguous, leading the model to an extra exposure to past instances, thus overfitting. The idea is reinforced by the CCA-GNN approach, where an opposite behavior is observed in most cases, i.e., since CCA-GNN does not implement a memory, higher numbers of neighbors usually lead to lower prediction errors.

\begin{table}[!htb]
\caption{Average Mean Squared Error and standard deviation over Canonical Cortical GNN and CCA-GNN considering clean audio reconstruction given noisy audio and visual inputs over two block layers network architecture.}
\begin{center}
\renewcommand{\arraystretch}{1.5}
\setlength{\tabcolsep}{6pt}
\resizebox{0.5\textwidth}{!}{
\begin{tabular}{c|c|c}
\hhline{-|-|-|}
\hhline{-|-|-|}
\hhline{-|-|-|}
{\cellcolor[HTML]{EFEFEF}{\textbf{Neighbors}}}& {\cellcolor[HTML]{EFEFEF}{\textbf{CCA GNN}}}& {\cellcolor[HTML]{EFEFEF}{\textbf{Cortical}}}\\ \hline
\textbf{3} & $0.0204\pm0.0048$ & $\bm{0.0188\pm0.0041}$\\ \hline
\textbf{5} & $0.0208\pm0.0049$ & $0.0200\pm0.0057$\\ \hline
\textbf{7} & $0.0214\pm0.0052$ & $0.0243\pm0.0067$\\ \hline
\textbf{10} & $0.0216\pm0.0052$ & $0.0282\pm0.0059$\\ \hline
\textbf{15} & $0.0216\pm0.0048$ & $0.0278\pm0.0049$\\ \hline
\textbf{20} & $0.0205\pm0.0045$ & $0.0266\pm0.0048$\\ \hline
\textbf{25} & $0.0200\pm0.0045$ & $0.0262\pm0.0056$\\ \hline
\textbf{30} & $0.0199\pm0.0045$ & $0.0246\pm0.0058$\\ \hline
\hhline{-|-|-|}
\hhline{-|-|-|}
\hhline{-|-|-|}
\end{tabular}}
\label{t.reconstruction_audioVisual_2}
\end{center}
\end{table}


Finally, Figure~\ref{f.reconstruction_convergence} depicts some examples of clean audio reconstruction for both models. Notice that the Canonical Cortical Graph Neural Networks with a reduced number of past frames, i.e., $k=3$, obtained almost perfect reconstructions of the clean signal in both cases, which reinforces the idea that the memory implementation replaces the necessity of longer temporal information, described by a higher number of neighbors. Figure~\ref{f.reconstruction_convergence}(b) shows that CCA-GNN is also capable of producing good representations when then the number of past frames is big enough, i.e., $k=30$, even though the reconstruction is not as good as the proposed method.

\begin{figure}[!htb]
  \centerline{
    \begin{tabular}{cc}
	\includegraphics[width=0.35\textwidth]{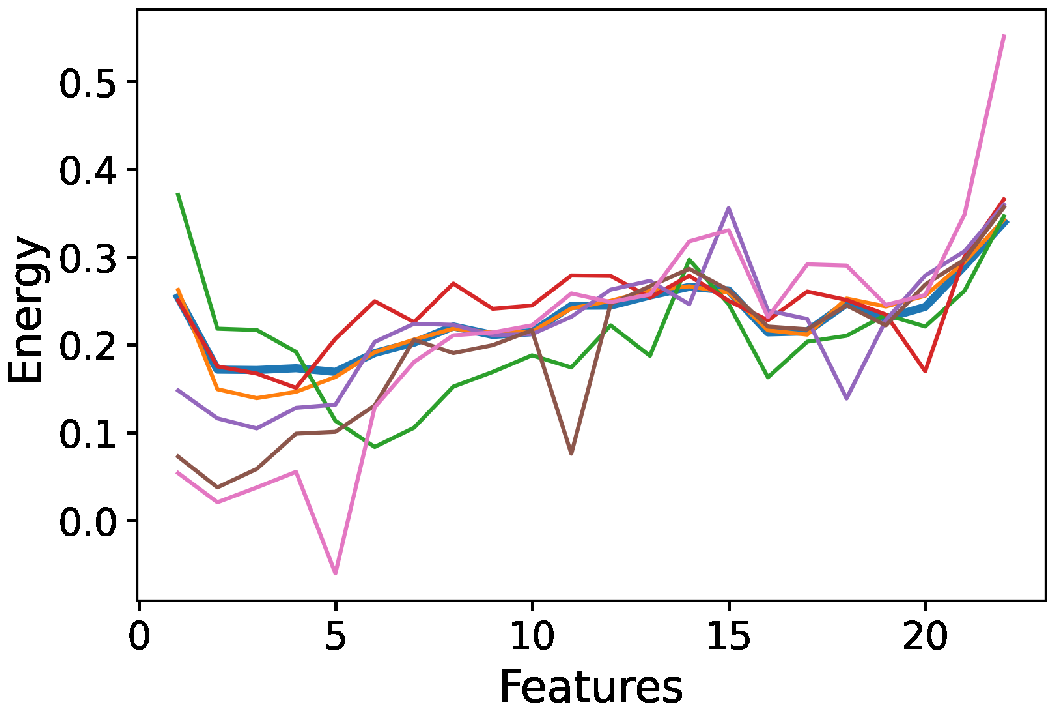} &
	\includegraphics[width=0.62\textwidth]{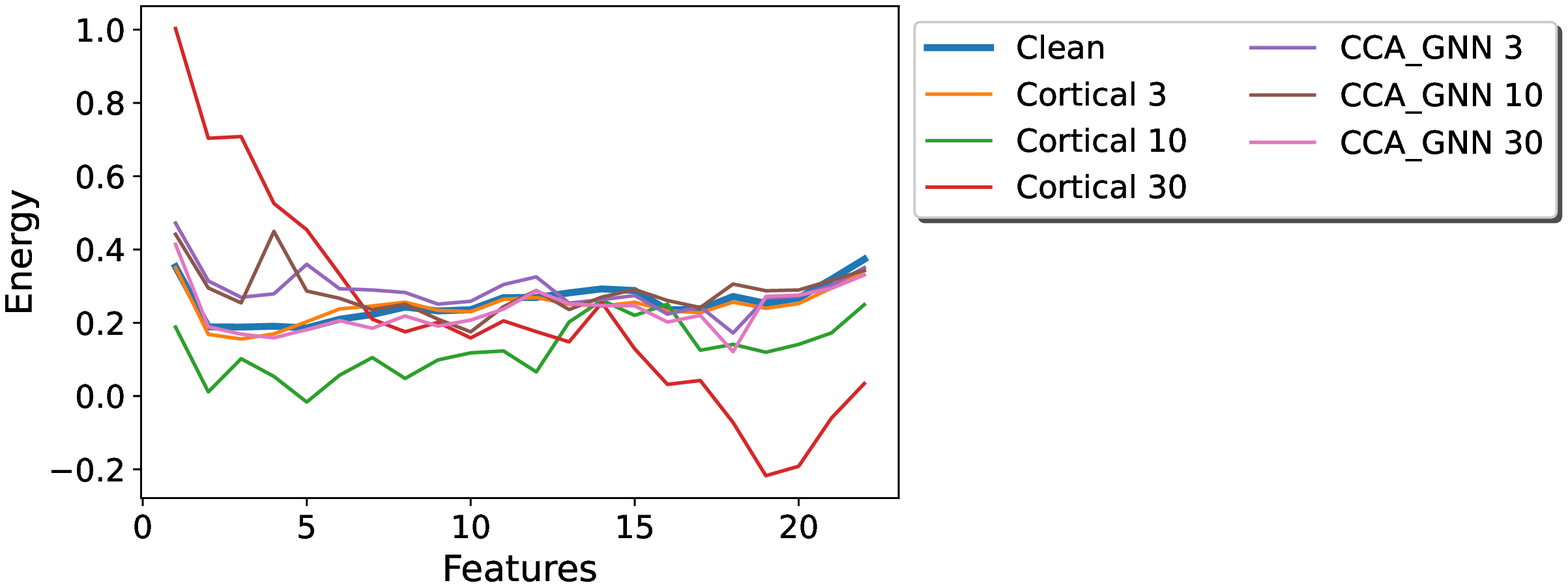} \\
	(a) & (b)
    \end{tabular}}
    \caption{Two examples of clean audio signal reconstruction considering the proposed approach (Cortical) and the baseline, i.e., CCA-GNN.}
  \label{f.reconstruction_convergence}
\end{figure}

\subsection{State-of-the-art Comparison}
\label{ss.stateOfArt}

Table~\ref{t.otherWorks} provides a comparison of the Canonical Cortical Graph Neural Network against the state-of-the-art results reported in recent works using the AV ChiME3 dataset and multimodal approaches for audio-visual speech enhancement, i.e., CCA-GNN and CCA Multilayer Perceptron (CCA-MLP)~\cite{passos2022}, a Long-Short Term Memory (LSTM)~\cite{adeel2020novel} and a Multilayer Perceptron (MLP)~\cite{adeel2019lip} based approaches, as well as a  canonical correlation-based short-time objective intelligibility deep learning (CC-STOI DL)~\cite{hussain2022novel} method. Notice the proposed approach provided the most accurate results, outperformed all the compared tecniques.

\begin{table}[!htb]
\caption{Comparison of the Canonical Cortical Graph Neural Network against state-of-the-art results reported in recent approaches for audio-visual speech reconstruction in terms of ean squared error.}
\begin{center}
\renewcommand{\arraystretch}{1.5}
\setlength{\tabcolsep}{6pt}
\resizebox{0.5\textwidth}{!}{
\begin{tabular}{c|c}
\hhline{-|-|}
\hhline{-|-|}
\hhline{-|-|}
{\cellcolor[HTML]{EFEFEF}{\textbf{Technique}}}&  {\cellcolor[HTML]{EFEFEF}{\textbf{Best value}}}\\ \hline
\textbf{Cortical (ours)} &  $0.0188$\\ \hline
\textbf{CCA-GNN} &  $0.0199$\\ \hline
\textbf{CCA-MLP~\cite{passos2022}} & $0.0189$\\ \hline
\textbf{MLP~\cite{adeel2019lip}} & $0.0200$\\ \hline
\textbf{LSTM~\cite{adeel2020novel}} &  $0.0780$\\ \hline
\textbf{CC-STOI DL~\cite{hussain2022novel}} & $1.2770$\\ \hline
\hhline{-|-|}
\hhline{-|-|}
\hhline{-|-|}
\end{tabular}}
\label{t.otherWorks}
\end{center}
\end{table}

\subsection{Neuronal Activation Analysis}
\label{ss.neuronActivation}

\begin{figure}[!htb]
  \centerline{
    \begin{tabular}{cc}
	\includegraphics[width=.42\textwidth]{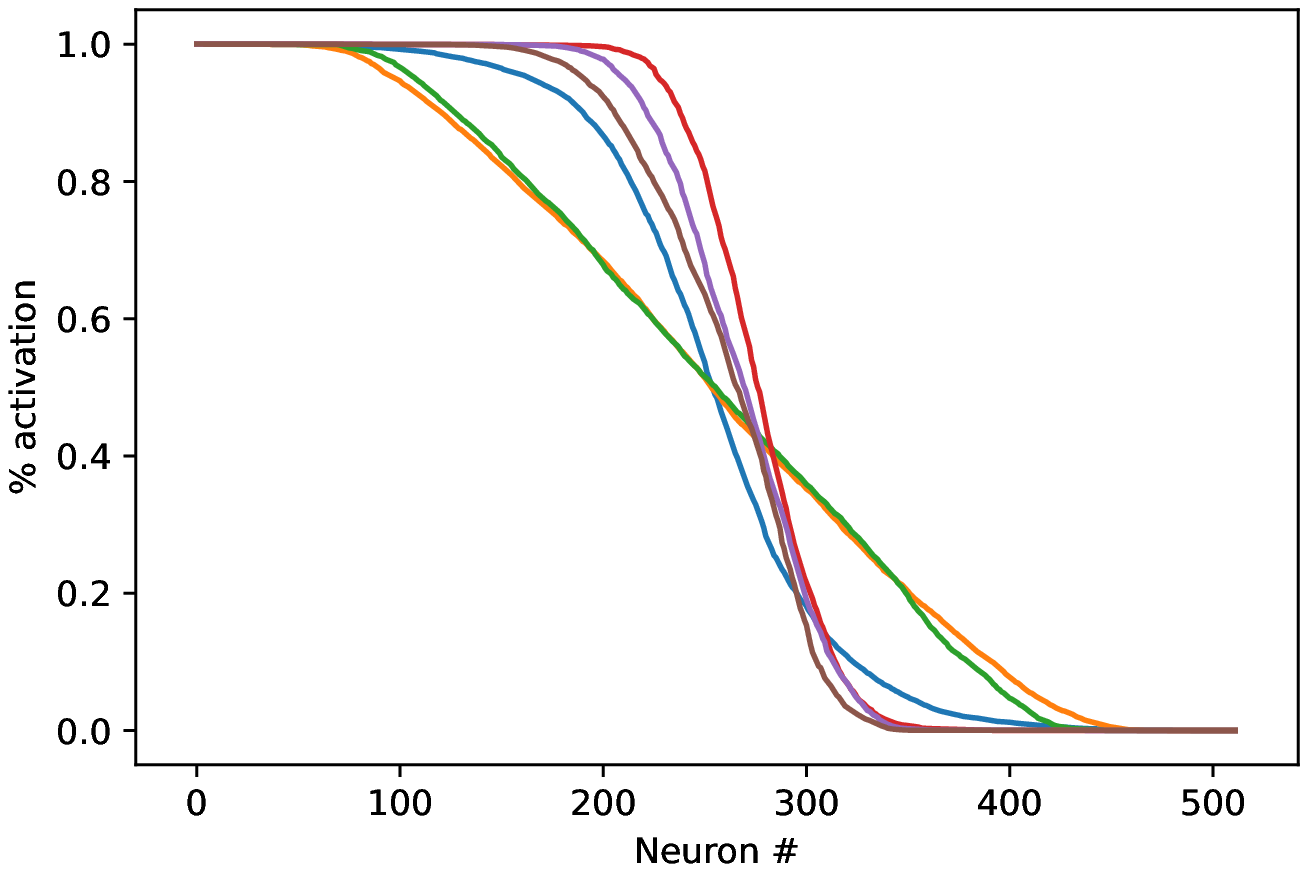} &
	\includegraphics[width=.58\textwidth]{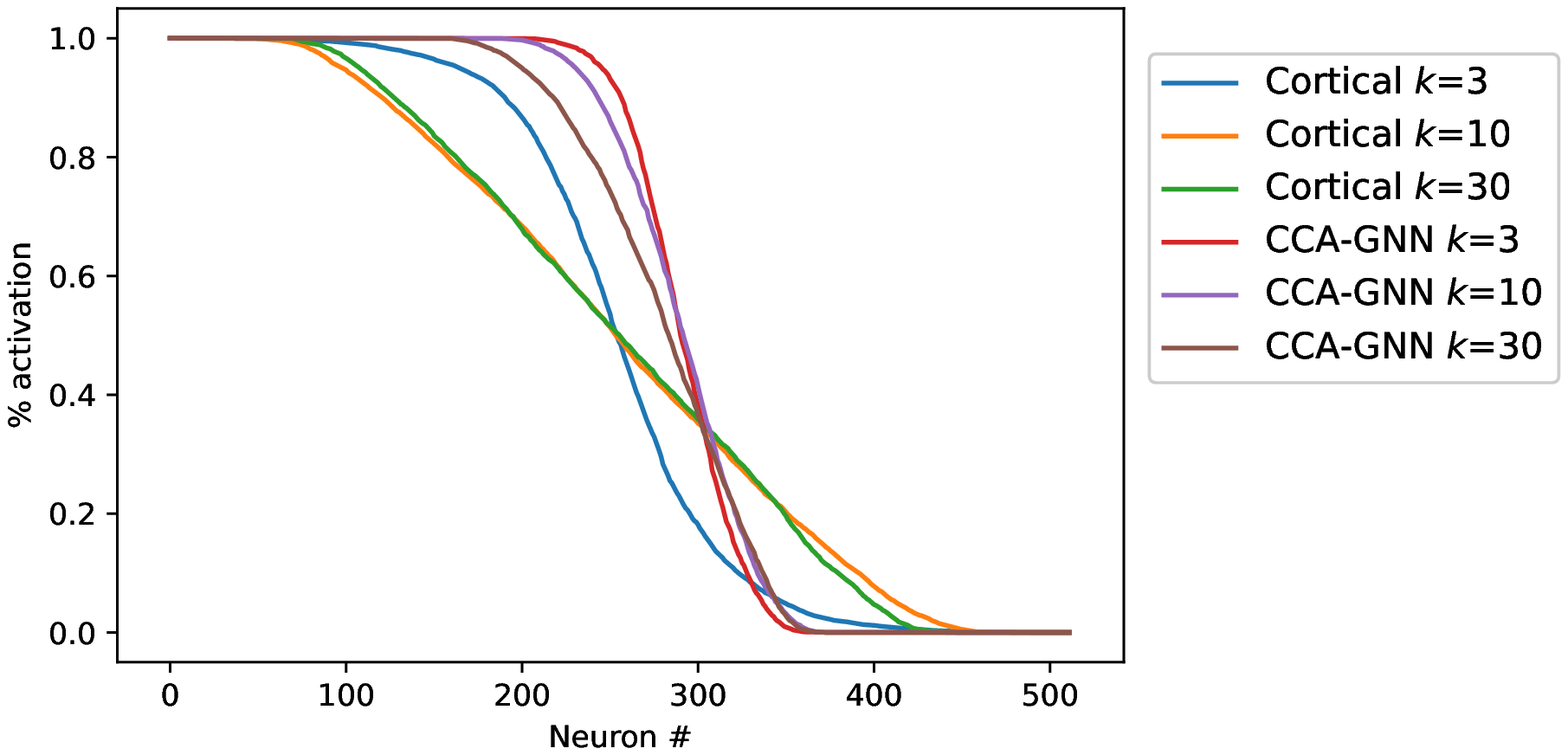} \\
	(a) & (b)
    \end{tabular}}
    \caption{Neuron activation rate considers the multimodal architecture's hidden block over (a) noisy audio and (b) visual channels.}
  \label{f.activation_2}
\end{figure}

The neuronal activation rate analysis is fundamental for future audio-visual hearing aids since the metric reflects the model's energy efficiency, a critical feature considering energy-constrained environments like hearing aids and embedded devices. In this context, reducing the firing rate directly implies a reduction in energy consumption. Figure~\ref{f.activation_2} depicts the neuron activation rate over the intermediate block layers concerning the audio and visual channels. Notice that both audio and video channels obtained pretty similar values concerning the Cortical model, suggesting that the two tracks contribute almost equally to the results.

A numerical representation is presented in Table~\ref{t.activation_2}, which comprises a more complete set of neighbors (past frames), i.e., $k=[3,5,7,10,15,20,25]$. The results reinforce the idea that the proposed model relies almost equally upon both noisy audio and visual modalities for the final output since their areas under the curve are practically the same for all cases, except for $k=10$, with an irrelevant difference, i.e., $256$ and $255$ for noisy audio and visual, respectively. The results also show that the proposed model outperforms the baseline in the context of neuronal activation and, consequently, energy consumption in every possible scenario, showing itself as a suitable approach for speech enhancement in future audio-visual hearing aid devices considering both accuracy and energy performance.

\begin{table}[!htb]
\caption{Area under the curve considering  the multimodal architecture's hidden block.}
\begin{center}
\renewcommand{\arraystretch}{1.5}
\setlength{\tabcolsep}{6pt}
\resizebox{0.6\textwidth}{!}{
\begin{tabular}{c|c|c|c|c}
\hhline{-|-|-|-|-|}
\hhline{-|-|-|-|-|}
\hhline{-|-|-|-|-|}
\cellcolor[HTML]{EFEFEF}&\multicolumn{2}{c|}{\cellcolor[HTML]{EFEFEF}\textbf{Audio}}&\multicolumn{2}{c}{\cellcolor[HTML]{EFEFEF}\textbf{Visual}}\\\hhline{~----}
\multirow{-2}{*}{\cellcolor[HTML]{EFEFEF}{\textbf{Neighbors}}} & {\cellcolor[HTML]{EFEFEF}{\textbf{Cortical}}}& {\cellcolor[HTML]{EFEFEF}{\textbf{CCA-GNN}}}& {\cellcolor[HTML]{EFEFEF}{\textbf{Cortical}}}& {\cellcolor[HTML]{EFEFEF}{\textbf{CCA-GNN}}}\\ \hline
$3$ & $255$ & $276$  & $255$& $291$    \\ \cline{1-5}
$5$ &   $253$ & $265$&  $253$& $288$   \\ \cline{1-5}
$7$ &  $256$ & $266$&  $256$& $288$  \\ \hline
$10$ &  $256$ & $268$& $255$& $290$    \\ \hline
$15$ &  $255$ & $268$&  $255$& $288$    \\ \hline
$20$ &  $258$ & $267$&  $258$& $285$    \\ \hline
$25$ &  $251$ & $267$&  $251$& $288$    \\ \hline
$30$ &  $255$ & $260$&  $255$& $280$    \\ \hline
\hhline{-|-|-|-|-|}
\hhline{-|-|-|-|-|}
\hhline{-|-|-|-|-|}
\end{tabular}}
\label{t.activation_2}
\end{center}
\end{table}

\section{Conclusion}
\label{s.conclusion}

This paper proposed a novel self-supervised method for multimodal correlated feature extraction through canonical correlation analysis maximization. The proposed model comprises a block-based neural network, where each block comprises two Graph Neural Networks layers, i.e., one for noisy audio and the other for the visual features, a memory, and a set of operations to filter, insert, delete, and modulate the input signals. Such operations are inspired in recent discoveries related to cortical cells and their interactions.

Experiments were conducted over the AV ChiME3 dataset, designed for the task of multimodal clean audio reconstruction considering noisy audio and clean visual instances, compared the proposed approach against the CCA-GNN, a similar state-of-the-art method proposed recently for the task. Results show that the proposed Canonical Cortical GNN provides more coherent and better-quality features, reaching higher values of canonical correlation analysis. The proposed approach also obtained more accurate reconstructions, generating cleaner reconstructions. Moreover, the model delivers higher efficiency in terms of energy, evaluated by neurons' firing rate. Finally, it also showed itself to be less dependent on a more extended prior-frame sequence, i.e., high values of $k$, since the memory can store and track temporal information.

Regarding future work, we aim to extend a similar cortical-based architecture to Convolutional Neural Networks and applications to two-dimensional data. We also aim to implement the model on chips for training and inference for possible future implementation in hearing aid devices for AV speech enhancement.

\bibliographystyle{splncs04}
\bibliography{refs}

\begin{thebibliography}{10}
\providecommand{\url}[1]{\texttt{#1}}
\providecommand{\urlprefix}{URL }
\providecommand{\doi}[1]{https://doi.org/#1}

\bibitem{abel2016data}
Abel, A., Marxer, R., Barker, J., Watt, R., Whitmer, B., Derleth, P., Hussain,
  A.: A data driven approach to audiovisual speech mapping. In: International
  Conference on Brain Inspired Cognitive Systems. pp. 331--342. Springer (2016)

\bibitem{adeel2018real}
Adeel, A., Ahmad, J., Hussain, A.: Real-time lightweight chaotic encryption for
  5g iot enabled lip-reading driven secure hearing-aid. arXiv preprint
  arXiv:1809.04966  (2018)

\bibitem{adeel2020novel}
Adeel, A., Ahmad, J., Larijani, H., Hussain, A.: A novel real-time, lightweight
  chaotic-encryption scheme for next-generation audio-visual hearing aids.
  Cognitive Computation  \textbf{12}(3),  589--601 (2020)

\bibitem{adeel2020contextual}
Adeel, A., Gogate, M., Hussain, A.: Contextual deep learning-based audio-visual
  switching for speech enhancement in real-world environments. Information
  Fusion  \textbf{59},  163--170 (2020)

\bibitem{adeel2019lip}
Adeel, A., Gogate, M., Hussain, A., Whitmer, W.M.: Lip-reading driven deep
  learning approach for speech enhancement. IEEE Transactions on Emerging
  Topics in Computational Intelligence  (2019)

\bibitem{barker2015third}
Barker, J., Marxer, R., Vincent, E., Watanabe, S.: The third ‘chime’speech
  separation and recognition challenge: Dataset, task and baselines. In: 2015
  IEEE Workshop on Automatic Speech Recognition and Understanding (ASRU). pp.
  504--511. IEEE (2015)

\bibitem{benesty2011speech}
Benesty, J., Chen, J., Habets, E.A.: Speech enhancement in the STFT domain.
  Springer Science \& Business Media (2011)

\bibitem{bokhari2013multimodal}
Bokhari, M.U., Hasan, F.: Multimodal information retrieval: Challenges and
  future trends. International Journal of Computer Applications
  \textbf{74}(14) (2013)

\bibitem{canonicalCortical}
Capone, F., Paolucci, M., Assenza, F., Brunelli, N., Ricci, L., Florio, L.,
  Di~Lazzaro, V.: Canonical cortical circuits: current evidence and theoretical
  implications. Neuroscience and Neuroeconomics  \textbf{5}, ~1--8 (2016)

\bibitem{chang2018scalable}
Chang, X., Xiang, T., Hospedales, T.M.: Scalable and effective deep cca via
  soft decorrelation. In: Proceedings of the IEEE Conference on Computer Vision
  and Pattern Recognition. pp. 1488--1497 (2018)

\bibitem{cooke2006audio}
Cooke, M., Barker, J., Cunningham, S., Shao, X.: An audio-visual corpus for
  speech perception and automatic speech recognition. The Journal of the
  Acoustical Society of America  \textbf{120}(5),  2421--2424 (2006)

\bibitem{da2020critical}
da~Costa, K.A., Papa, J.P., Passos, L.A., Colombo, D., Del~Ser, J., Muhammad,
  K., de~Albuquerque, V.H.C.: A critical literature survey and prospects on
  tampering and anomaly detection in image data. Applied Soft Computing p.
  106727 (2020)

\bibitem{gogate2020cochleanet}
Gogate, M., Dashtipour, K., Adeel, A., Hussain, A.: Cochleanet: A robust
  language-independent audio-visual model for real-time speech enhancement.
  Information Fusion  \textbf{63},  273--285 (2020)

\bibitem{canonicalLaminar}
Grossberg, S.: A canonical laminar neocortical circuit whose bottom-up,
  horizontal, and top-down pathways control attention, learning, and
  prediction. Frontiers in Systems Neuroscience  \textbf{15} (2021)

\bibitem{helvik2006psychological}
Helvik, A.S., Jacobsen, G., Hallberg, L.R.: Psychological well-being of adults
  with acquired hearing impairment. Disability and rehabilitation
  \textbf{28}(9),  535--545 (2006)

\bibitem{huang2021hearing}
Huang, A.R., Deal, J.A., Rebok, G.W., Pinto, J.M., Waite, L., Lin, F.R.:
  Hearing impairment and loneliness in older adults in the united states.
  Journal of Applied Gerontology  \textbf{40}(10),  1366--1371 (2021)

\bibitem{hussain2022novel}
Hussain, T., Diyan, M., Gogate, M., Dashtipour, K., Adeel, A., Tsao, Y.,
  Hussain, A.: A novel speech intelligibility enhancement model based on
  canonicalcorrelation and deep learning. arXiv preprint arXiv:2202.05756
  (2022)

\bibitem{kording2001supervised}
K{\"o}rding, K.P., K{\"o}nig, P.: Supervised and unsupervised learning with two
  sites of synaptic integration. Journal of computational neuroscience
  \textbf{11}(3),  207--215 (2001)

\bibitem{kramer1995factors}
Kramer, S.E., Kapteyn, T.S., Festen, J.M., Tobi, H.: Factors in subjective
  hearing disability. Audiology  \textbf{34}(6),  311--320 (1995)

\bibitem{kumar2022deep}
Kumar, L.A., Renuka, D.K., Rose, S.L., Wartana, I.M., et~al.: Deep learning
  based assistive technology on audio visual speech recognition for hearing
  impaired. International Journal of Cognitive Computing in Engineering
  \textbf{3},  24--30 (2022)

\bibitem{ngiam2011multimodal}
Ngiam, J., Khosla, A., Kim, M., Nam, J., Lee, H., Ng, A.Y.: Multimodal deep
  learning. In: ICML (2011)

\bibitem{noble1998self}
Noble, W.: Self-assessment of hearing and related function. Wiley-Blackwell
  (1998)

\bibitem{world2021hearing}
Organization, W.H., et~al.: Hearing screening: considerations for
  implementation  (2021)

\bibitem{passos2022multimodal}
Passos, L.A., Khubaib, A., Raza, M., Adeel, A.: Multimodal speech enhancement
  using burst propagation. arXiv preprint arXiv:2209.03275  (2022)

\bibitem{passos2022}
Passos, L.A., Papa, J.P., Del~Ser, J., Hussain, A., Adeel, A.: Multimodal
  audio-visual information fusion using canonical-correlated graph neural
  network for energy-efficient speech enhancement. Information Fusion  (2022)

\bibitem{payeur2021burst}
Payeur, A., Guerguiev, J., Zenke, F., Richards, B.A., Naud, R.: Burst-dependent
  synaptic plasticity can coordinate learning in hierarchical circuits. Nature
  neuroscience  \textbf{24}(7),  1010--1019 (2021)

\bibitem{ross2008incremental}
Ross, D.A., Lim, J., Lin, R.S., Yang, M.H.: Incremental learning for robust
  visual tracking. International journal of computer vision  \textbf{77}(1),
  125--141 (2008)

\bibitem{santos2021ddipnet}
Santos, D.F., Pires, R.G., Passos, L.A., Papa, J.P.: Ddipnet and ddipnet+:
  Discriminant deep image prior networks for remote sensing image
  classification. In: 2021 IEEE International Geoscience and Remote Sensing
  Symposium IGARSS. pp. 2843--2846. IEEE (2021)

\bibitem{de2021computer}
de~Souza, R.W., Silva, D.S., Passos, L.A., Roder, M., Santana, M.C., Pinheiro,
  P.R., de~Albuquerque, V.H.C.: Computer-assisted parkinson's disease diagnosis
  using fuzzy optimum-path forest and restricted boltzmann machines. Computers
  in Biology and Medicine  \textbf{131},  104260 (2021)

\bibitem{thakoor2021bootstrapped}
Thakoor, S., Tallec, C., Azar, M.G., Munos, R., Veli{\v{c}}kovi{\'c}, P.,
  Valko, M.: Bootstrapped representation learning on graphs. arXiv preprint
  arXiv:2102.06514  (2021)

\bibitem{tian2021understanding}
Tian, Y., Chen, X., Ganguli, S.: Understanding self-supervised learning
  dynamics without contrastive pairs. arXiv preprint arXiv:2102.06810  (2021)

\bibitem{viola2001rapid}
Viola, P., Jones, M.: Rapid object detection using a boosted cascade of simple
  features. In: Proceedings of the 2001 IEEE computer society conference on
  computer vision and pattern recognition. CVPR 2001. vol.~1, pp.~I--I. Ieee
  (2001)

\bibitem{Wilcoxon:45}
Wilcoxon, F.: Individual comparisons by ranking methods. Biometrics Bulletin
  \textbf{1}(6),  80--83 (1945). \doi{https://doi.org/10.2307/3001968}

\bibitem{xu2014regression}
Xu, Y., Du, J., Dai, L.R., Lee, C.H.: A regression approach to speech
  enhancement based on deep neural networks. IEEE/ACM Transactions on Audio,
  Speech, and Language Processing  \textbf{23}(1),  7--19 (2014)

\bibitem{yu2022setransformer}
Yu, W., Zhou, J., Wang, H., Tao, L.: Setransformer: speech enhancement
  transformer. Cognitive Computation  \textbf{14}(3),  1152--1158 (2022)

\bibitem{zeng2022small}
Zeng, N., Wu, P., Wang, Z., Li, H., Liu, W., Liu, X.: A small-sized object
  detection oriented multi-scale feature fusion approach with application to
  defect detection. IEEE Transactions on Instrumentation and Measurement
  \textbf{71},  1--14 (2022)

\bibitem{zhang2021canonical}
Zhang, H., Wu, Q., Yan, J., Wipf, D., Philip, S.Y.: From canonical correlation
  analysis to self-supervised graph neural networks. In: Thirty-Fifth
  Conference on Neural Information Processing Systems (2021)

\bibitem{zhu2020deep}
Zhu, Y., Xu, Y., Yu, F., Liu, Q., Wu, S., Wang, L.: Deep graph contrastive
  representation learning. arXiv preprint arXiv:2006.04131  (2020)

\end{thebibliography}
\end{document}